# Gate-induced insulating state in bilayer graphene devices

Jeroen B. Oostinga, Hubert B. Heersche, Xinglan Liu, Alberto F. Morpurgo and Lieven M. K. Vandersypen

*Kavli Institute of Nanoscience, Delft University of Technology, PO Box 5046, 2600 GA, Delft, The Netherlands.*

**The potential of graphene-based materials consisting of one or a few layers of graphite for integrated electronics originates from the large room-temperature carrier mobility in these systems (~10 , 000 cm2/Vs). However, the realization of electronic devices such as field-effect transistors will require controlling and even switching off the electrical conductivity by means of gate electrodes, which is made difficult by the absence of a bandgap in the intrinsic material. Here, we demonstrate the controlled induction of an insulating state - with large suppression of the conductivity - in bilayer graphene, by using a double-gate device configuration that allows an electric field to be applied perpendicular to the plane. The dependence of the resistance on temperature and electric field, and the absence of any effect in a single-layer device, strongly suggest that the gate-induced insulating state originates from the recently predicted opening of a bandgap between valence and conduction bands.**

Graphene systems, consisting of one or a few crystalline monolayers of carbon atoms, stand out because of their unusual electronic properties and of their potential for applications in nanoelectronics [1-5]. Carrier mobility values as high as 10000 $cm^2/Vs$ at room temperature -ten times higher than in silicon- are routinely obtained in these materials, without the need for sophisticated preparation techniques [1]. Both the high mobility and the envisioned possibility of low-cost mass-production provide a strong drive to explore the use of graphene for future high-speed integrated electronic circuits. In order to develop such "graphene-based electronics", however, several problems need to be overcome. Perhaps the most important obstacle is the absence of an energy gap separating the valence and conduction band of graphene –graphene is a zero-gap semiconductor [6]. As a consequence, electrical conduction cannot be switched off using control voltages [7], which is essential for the operation of conventional transistors. It was recently shown that conduction can be switched off by patterning single-layer graphene into narrow ribbons [8]. Here, we demonstrate that we can produce an insulating state and switch off electrical conduction in a bilayer graphene device, simply by applying control voltages to two on-chip gate electrodes.

Our strategy is motivated by recent theoretical work that discusses how a band-gap can be opened in single- and bi-layer graphene [9,10]. To understand the physical mechanisms underlying these predictions, we consider the basic electronic properties of graphene-based materials in some detail. Monolayer graphene has a honeycomb lattice structure with a unit cell consisting of two atoms – normally referred to as A and B atoms (figure 1a). The Hamiltonian that describes the electronic properties of graphene near the Fermi level can be approximated as [6,11]

$$H = \begin{pmatrix} \Delta & \hbar v_F \left(k_x - ik_y\right) \\ \hbar v_F \left(k_x + ik_y\right) & -\Delta \end{pmatrix} \qquad (1),$$

with $k$ the momentum, and $v_F$ the Fermi velocity. This operator acts on spinors $\psi = \begin{pmatrix} \phi_A \\ \phi_B \end{pmatrix}$, with $\phi_A$ and $\phi_B$ the amplitudes of the wavefunction on sublattices A and B, and $\Delta$ is the on-site energy difference between the two sublattices. Normally $\Delta = 0$ and this Hamiltonian results in the Dirac-like linear dispersion relation $E = \pm\hbar v_F |k|$ (figure 1a). The positive and negative solutions, which correspond to conduction and valence bands respectively, meet at $k = 0$, implying the absence of a band-gap. In order to open a gap, the inversion symmetry in the graphene plane must be broken by making $\Delta \neq 0$. In this case, the low-energy Hamiltonian (1) leads to a gapped dispersion relation $E(k) = \pm\sqrt{\Delta^2 + (\hbar v_F k)^2}$. This inversion symmetry breaking can in principle be implemented experimentally. For instance, one can envision placing graphene onto a boron-nitride (BN) substrate that has the same honeycomb lattice structure and comparable lattice spacing, so that the A and B atoms experience different on-site energies [9]. In practice, however, the technological challenges that need to be met to implement such a strategy are highly non-trivial.

In bilayer graphene, in contrast, a conceptually similar strategy is within technological reach. Bilayer graphene consists of two monolayers stacked as in natural graphite (figure 1b). This so-called Bernal stacking yields a unit cell of four atoms (one atom of each of the sublattices A1, B1, A2 and B2) resulting in four electronic bands. Only two of these bands are relevant at low energy; they can be described by the effective Hamiltonian [12]

$$H = \begin{pmatrix} \Delta & -\frac{\hbar^2}{2m}\left(k_x - ik_y\right)^2 \\ -\frac{\hbar^2}{2m}\left(k_x + ik_y\right)^2 & -\Delta \end{pmatrix} \qquad (2).$$

This operator has a structure similar to that of equation (1) and, as for the monolayer, it also leads to a spectrum with zero-gap between valence and conduction band when $\Delta = 0$, but now with a quadratic dispersion relation ($E = \pm\hbar^2 k^2 / 2m$; figure 1b). Furthermore, and essential for our purposes, the operator acts on spinors $\psi = \begin{pmatrix} \phi_{A1} \\ \phi_{B2} \end{pmatrix}$ which contain the amplitude of the wavefunction on atoms A1 and B2 *that are located in the two different layers*. This makes it possible to control the difference between the on-site energy of A1 and B2 electrostatically, simply by applying a sufficiently strong electric field *E* perpendicular to the carbon atom planes. In the presence of such an electric field, a gap of size $2\Delta$ opens between conduction and valence band (figure 1c) [10,12,13]. Indeed, a band-gap originating from this mechanism has been recently observed in angle-resolved photoemission spectroscopy experiments on a chemically doped graphene bilayer, in which the electric field is associated with the charge transfer from the dopants to the carbon atoms [14] (see also [15]). Here, we use a double-gate device configuration to impose a perpendicular electric field onto a graphene bilayer, which allows us to demonstrate the controlled transition from a zero-gap semiconductor to an insulator, by simply adjusting the voltages applied to the two gate electrodes.

Figure 1d shows the device configuration that we investigate. It consists of single or double graphene layers sandwiched in between two gate electrodes, and connected to metallic leads. These double-gated structures enable simultaneous and independent control of the charge density in the system (i.e., the position of the Fermi level) and of the electric field perpendicular to the graphene layer. In a single layer, the presence of a perpendicular field is not expected to affect the transport properties: the conductivity of the device should never become smaller than a minimum value of the order of $4e^2/h$, irrespective of the applied gate voltages [2]. In a bilayer, on the contrary, a large applied field results in a different electrostatic potential in the two layers, which, according to theory, should cause a band-gap to open. If the Fermi level is maintained in the gap (i.e. the device is operated near the charge neutrality point), this should result in an insulating temperature dependence of the conductivity, dropping to well below $4e^2/h$ at low temperature. A unique signature of this effect is that the decrease in conductivity with lowering temperature becomes more pronounced for larger applied electric field values. This possibility to controllably induce an insulating state, which is crucial for switching devices, was missing in earlier experiments on graphene bilayers [14,15] where the gap and the carrier density could not be gate-controlled independently.

The fabrication of double-gated graphene devices is similar to what has been described elsewhere [1], and relies on micromechanical cleaving of natural graphite. The flakes used in the experiments were selected under an optical microscope and identified as single- and double-layer graphene, respectively, based on their optical contrast (see supplementary material; a similar method is used as previously demonstrated in [16,17]). Contact to the flakes was made by means of electron-beam lithography, electron-beam evaporation of a Ti/Au bilayer (10/50 nm), and lift-off. The top-gate insulating layer and electrodes were defined subsequently, by e-beam deposition of a $SiO_2$ (15 nm) followed by deposition of a Ti/Au bilayer (6.5/40 nm), without breaking the vacuum. The comparison between single- and double-layer graphene devices is useful not only to illustrate the profound difference between them and to identify the mechanism responsible for the gate-induced insulating state, but also to rule out possible spurious effects originating from the device fabrication (e.g., damage to the graphene layers or disorder introduced by the deposition of the $SiO_2$ gate dielectric).

We now proceed to discuss the systematic transport measurements that we have performed, starting with the single-layer device shown in figure 2a. Figures 2b and 2c show the resistance measured as a function of the voltage applied to one of the gates, with the other gate at a constant potential as indicated (we extract a carrier mobility of ~3000 $cm^2/Vs$, similar to the mobility of an ungated device on the same flake). Irrespective of which gate voltage is kept constant, we always observe a peak in resistance characteristic of the behaviour of few-layer graphene and hereafter referred to as the "charge-neutrality (CN) peak" (to be precise, we are measuring a device comprising regions of different carrier density, n1-n2-n1; at the resistance maximum, n2=0 only). The position of the CN peak when sweeping one gate shifts linearly with the voltage applied to the other gate. Irrespective of the gate voltage configuration, the height of the resistance peak remains approximately constant. From the top-gate dimensions, we can estimate a minimum conductivity value close to $4e^2/h$ (also the part of the flake that is not covered by the top gate contributes to the resistance, but given the dimensions, this increases the conductivity estimate by at most a factor of 1.4). This is typical of graphene at the charge neutrality point, which indicates that the device fabrication and the deposition of the top-gate dielectric have not resulted in substantial damage to the material. Note also that depending on the values of the voltages applied to both gates *pn*-junctions are formed near the interfaces between the region covered by the top-gate and both uncovered regions [18]. Such *pn*-junctions may be the origin of the weak asymmetry seen in many of the gate sweeps in figures 2b and 2c. However, near the charge-neutrality point, we expect *pn*-junctions to give only a small contribution to the measured resistance. Finally, figure 2d shows that the gate voltage dependence of the resistance is not affected by varying the temperature between 4 and 50 K, apart from reproducible conductance fluctuations that increase in magnitude as the temperature is lowered. These observations are consistent with the expected behaviour of electrical transport through graphene monolayers [2, 18-20].

The behaviour of the double-gated graphene bilayer device (figure 3a) is strikingly different. Figures 3b and 3c show the (square) resistance of bilayer graphene as a function of the back-gate and top-gate voltages (the carrier mobility is ~1000 $cm^2$/Vs, again similar to an ungated device on the same flake). Similarly to the monolayer, the position of the CN peak shifts linearly with the respective gate voltages. Contrary to the monolayer, the CN peaks are nearly perfectly symmetric, ruling out the possibility that the formation of *pn*-junctions gives a dominant contribution to the measured resistance. More importantly, the maximum resistance value now depends on the configuration of gate voltages. Specifically, when the voltage applied to both gates is close to 0 V, the height of the CN peak corresponds to a conductivity of the order of $4e^2/h$, which is typical for zero-gap bilayers [5] (again we rely on the fact that near the CN peak the region under the top-gate gives the largest contribution to the resistance). However, as the top- and back-gate are biased with opposite voltages of increasing magnitude, the height of the CN peak exhibits a pronounced rise. Also the temperature dependence observed in the bilayer device is markedly different from that measured in the single-layer device (figure 3d). For small gate voltages, the resistance near the CN peak is essentially temperature independent, characteristic of a zero-gap semiconductor. When the difference in top- and back-gate voltage is increased, however, the maximum resistance value also increases as the temperature is lowered. The observation of a conductivity much smaller than $4e^2/h$ exhibiting an insulating temperature dependence for oppositely biased gate electrodes is what we would expect qualitatively in a bilayer graphene device.

In order to confirm that in the double-gated bilayer device large differences in voltage between top- and back-gate do lead to an insulating state, we have performed measurements in a dilution refrigerator, in the temperature range between 50 mK and 1.2 K, where the increase in resistance with lowering temperature should be more pronounced. Indeed, figure 4a (note the logarithmic scale) shows that when top- and back-gate are biased asymmetrically, a very strong temperature dependence of the square resistance is observed near the CN peak, reaching values between 10 and 100 MΩ at 55 mK. This is in stark contrast to the case of small gate voltages (figure 4a, inset), for which a temperature independent resistance near the CN peak -corresponding to a conductivity of approximately $4e^2/h$ - persists down to the lowest temperature. The full dependence of the square resistance measured at 50 mK as function of the voltage applied to both gate electrodes is shown in figure 4b, from which the very fast increase in resistance near the CN peak with increasing the electric field applied perpendicular to the layer is apparent. According to the expectations, the region of high resistance scales linearly with both top- and back-gate voltage (see white dotted line in the inset of figure 4b) as it is required to maintain charge neutrality in the graphene bilayer. Additionally, we have also measured the *I-V* characteristics of the device for different top- and back-gate voltage configurations (figure 4c) and observed that they evolve from exhibiting a linear Ohmic behaviour far from the CN peak, to a pronounced non-linear behaviour near the CN peak.

Finally, we discuss more quantitatively the insulating temperature dependence of the resistance that we observe for large oppositely biased gates. In an ideal defect-free insulator, thermally activated transport is expected, whereby the maximum resistance, *R*, varies with temperature as $R(T) \propto \exp\left(E_a / kT\right)$, with *k* Boltzmann's constant, and $E_a$ the activation energy, corresponding to half the band-gap. Our data, however, do not exhibit such a simple thermal activation behaviour (figure 5a). Below approximately $T$ = 5 K, and at the highest applied electric fields, they are much better described by $R(T) \propto \exp\left(T_0 / T\right)^{1/3}$, as seen in figure 5b, with fitted values of $T_0$ of ~0.5-0.8 K (note that between 5 K and 55 K the resistance drops more rapidly with increasing temperature, but the range is too small to deduce an accurate value for the activation energy). Qualitatively, a $n$=1/3 exponent is expected for transport in two dimensions of non-interacting carriers via variable-range hopping in insulating materials where transport is mediated by localized impurity sites which are present inside a gap [21,22]. Such localized states have been predicted theoretically in the case of disordered bilayer graphene [23], but drawing quantitative conclusions as to the properties of these states, e.g. their density of states, spatial extension, etc., is not straightforward from our measurements and goes beyond the scope of this paper. For smaller applied perpendicular electric fields the fitted exponent becomes smaller than 1/3 and decreases towards zero (see inset of figure 5b), and the fitted value of $T_0$ also decreases. This indicates clearly that the insulating temperature dependence of the resistance that we measure becomes stronger when the applied perpendicular electric field is higher.

We conclude that the data unambiguously show the occurrence of an insulating state in bilayer graphene in the presence of a perpendicular electric field, which has not been reported earlier. Our observation that the insulating state occurs only in bilayers and not in monolayers, and that the increase in resistance with lowering temperature is more pronounced for larger values of the electric field applied perpendicular to the material, is in agreement with the predicted controlled opening of a band-gap. In contrast, these two very specific observations cannot be accounted for simply by an increase in the amount of disorder, for instance caused by the presence of the top-gate. Furthermore, as pointed out earlier, carrier mobilities were comparable in devices with and without top-gates, actually suggesting equal amounts of disorder. Although we cannot rule out that other mechanisms than the formation of a band-gap could lead to the same striking observations, we are not aware of other possible explanations.

It is clear from the experiments that a possible gap induced in the bilayer device is rather small. This is consistent with recent theoretical calculations [10,13], which, for zero carrier density and for electric field values of the order of those achieved in our experiments, predict a gap size below approximately 10 meV depending on, for instance, the way in which screening effects in the bilayer are modeled (note that in [14,15], measurements were done at very high charge density, where much larger gap sizes were expected and observed). By comparison, the energy scale of disorder can be estimated to be of the order of a few meV from measurements of the spin splitting in the Quantum Hall regime [24], and somewhat larger from scanning SET experiments [25]. Altogether, it appears that a gap of below 10 meV in conjunction with the presence of sub-gap states originating from disorder, can account for the observed $\exp(T^{-1/3})$ dependence of the resistance in the temperature range between 50 mK and 4.2 K. However, this does not well explain the steep temperature dependence between 4.2 K and 55 K. Possibly the gap is in fact significantly larger than predicted, or alternatively there may be other mechanisms contributing as well to the observed resistance increase in this n1-n2-n1 device.

The possibility to use double-gated structures to suppress the conductivity of bilayer graphene to values much lower than $4e^2/h$ represents an important proof-of-principle for the feasibility of future graphene-based electronic devices. Obviously the development of practical devices will require further innovations, which are needed to switch off electrical conduction at room temperature. Nevertheless, the operation of devices at cryogenic temperatures that we have demonstrated here will already enable new fundamental studies of quantum transport in bilayer graphene, through the fabrication of structures such as quantum point contacts based on split-gates and electrostatically tunable quantum dots.

**ACKNOWLEDGEMENTS**

We gratefully acknowledge P. Jarillo-Herrero for experimental help in the early stages of this work, L.P. Kouwenhoven for providing access to a dilution refrigerator, E. McCann, A.H. MacDonald and H. Min for useful discussions, and NWO, FOM and NanoNed for financial support.

Supplementary information accompanies this paper on www.nature.com/naturematerials.

## FIGURE CAPTIONS

**Figure 1 Band-gap in graphene devices.** Schematics of the lattice structure of monolayer graphene (**a**) and bilayer graphene (**b**). The green and red coloured lattice sites indicate the A (A1/A2) and B (B1/B2) atoms of monolayer (bilayer) graphene, respectively. The diagrams represent the calculated energy dispersion relations in the low energy regime, and show that monolayer and bilayer graphene are zero-gap semiconductors (for bilayer graphene a pair of higher energy bands is also present, not shown in the diagram). **c**, When an electric field ($E_{\perp}$) is applied perpendicular to the bilayer, a band-gap is opened in bilayer graphene, whose size ($2\Delta$) is tunable by the electric field. **d**, Schematics of a double-gated graphene device as used in our investigations. Both the Fermi level in the graphene (bi)layer and the perpendicular electric field are controllable by means of the voltages applied to the back-gate, $V_{bg}$, and to the top-gate, $V_{tg}$. We study the resistivity of the graphene (bi)layer as a function of both gate voltages by applying a current bias ($I$) and measuring the resulting voltage across the device, $V$. Note the different $SiO_2$ thicknesses of the dielectric layers for the top- and back-gates.

**Figure 2 Gate voltage and temperature dependence of transport through monolayer graphene. a**, Optical microscope images of a single layer flake (left) and of the double-gated device fabricated on this flake (right). The yellow lines indicate metal contacts to the flake, and the blue line corresponds to the top-gate. The schematics of the four-probe device configuration is shown as well. **b**, Resistance versus back-gate voltage measured for different fixed values of the top-gate voltage showing an approximately gate-voltage independent height of the CN peak (the right axis gives an estimation of the square resistance, neglecting the contributions from the region without the top-gate and any *pn*-junctions to the measured resistance). The aperiodic fluctuations present near the CN peak are reproducible and are due to quantum interference. The smaller additional peak (indicated by the arrow) is the CN peak originating from the part of the flake that is not covered by the top-gate. **c**, Resistance versus top-gate voltage measured for different fixed values of back-gate voltage. Again, the height of the CN peak is nearly gate voltage independent. The difference in the magnitude of the voltage applied to top- and back-gates (**b**, **c**) originates from the different $SiO_2$ thicknesses separating the two gates from the graphene flake. **d**, Temperature dependence of the resistance versus top-gate voltage, measured for two different back-gate voltages. Irrespective of the gate voltage configuration the height of the CN peak is independent of temperature in the range 4.7 – 52 K.

**Figure 3 Gate voltage and temperature dependent transport through bilayer graphene. a**, Optical microscope images of a double layer flake (left) and of the double-gated device fabricated on this flake (right). Yellow lines represent metal contacts and the blue line represents the top-gate electrode. The two-probe measurement configuration is shown in the schematic (the measured resistance thus includes the contact resistance which is smaller than ~250 Ω). **b**, Resistance versus back-gate voltage measured for different fixed values of the top-gate voltage (the right axis gives the square resistance, again assuming that the region under the top-gate dominates the measured resistance; this assumption is valid near the CN peak where the resistance of the region with top-gate is relatively large). The height of the CN peak systematically increases when both gates are biased with increasingly large opposite voltages. **c**, Resistance versus top-gate voltage measured for different fixed back-gate voltages showing a similar gate voltage dependence of the height of the CN peak. **d**, Temperature dependence of the resistance versus top-gate voltage measured for two different values of back-gate voltage. When the voltage difference between both gates is small, the height of the CN peak is not affected by temperature in the range 4.2 – 55 K. However, a clear temperature dependence is observed in this same range when both gates are biased with large opposite voltages.

**Figure 4 Gate-induced insulating state in the bilayer graphene device. a**, Square resistance as a function of top-gate voltage measured at different temperatures: $T$ = 55 mK (blue), $T$ = 270 mK (red), $T$ = 600 mK (green) and $T$ = 1200 mK (black) (the back-gate voltage is kept fixed at $V_{bg}$=+50 V). Here, the square resistance is plotted, because near the CN peak the resistance is almost completely dominated by the region under the top-gate. A pronounced temperature dependence is observed when top- and back-gate are biased asymmetrically; when both gates are symmetrically biased no temperature dependence is observed (see inset). **b**, Three-dimensional plot of the square resistance as a function of both top- and back-gate voltage at $T$ = 50 mK, showing a sharp rise of the height of the CN peak with electric field. The inset is a colour plot of the same data, showing that the position of the CN peak shifts linearly with both gate voltages. The dark-coloured region corresponds to voltage configurations where an insulating state is observed. **c**, *I*-*V* characteristics measured at different gate voltage configurations (the letters correspond to the letters in the inset of **b,** which indicate both gate voltages). The data in **a** and **b** were taken with lock-in detection, using a zero voltage bias with a 5 μV excitation voltage modulated at 17 Hz. The data in **c** was measured with a DC voltage bias. Note that the plotted voltage bias is corrected for the internal resistance (1.1 kΩ) of the current measurement unit used in the experiment.

**Figure 5 Thermally activated hopping transport in biased bilayer graphene. a**, Logarithm of the square resistance at the CN peak versus inverse temperature, for different perpendicular electric fields ($E_{\perp} = \left(V_{bg} - V_{tg}\right) / \left(d_{bg} + d_{tg}\right)$ with $d_{bg}$ and $d_{tg}$ the thicknesses of the back- and top-gate oxides). These plots clearly show sub-linear behaviour in the temperature range from 55 mK to 55 K (the dotted line is just a guide to the eye), implying that the data cannot be described by $R(T) \propto \exp\left(E_a / kT\right)$. **b**, Logarithm of the square resistance of the CN peak plotted as function of $T^{-1/3}$ for different perpendicular electric fields in the temperature range from 55 mK to 55 K.

The linear fits to the data (solid lines) show that, at the highest fields, the data is well described by $R(T) \propto \exp\left(T_0 / T\right)^{1/3}$. At lower fields, the fitted exponent, *n*, is smaller than 1/3 (see inset; the error bars reflect the standard deviation of the fitted values).

# Figure 1

**Oostinga *et al.***

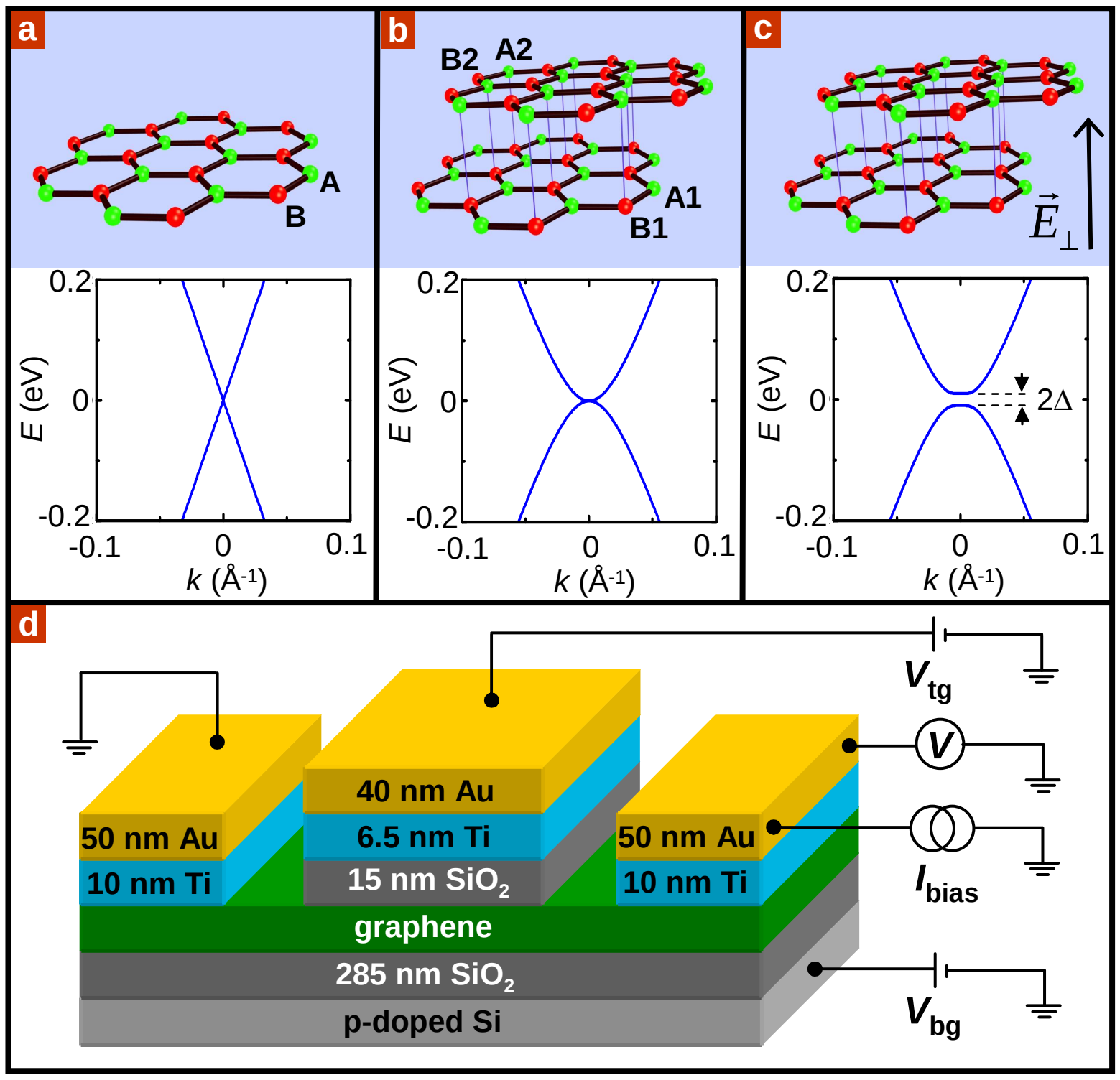

# Figure 2

**Oostinga *et al.***

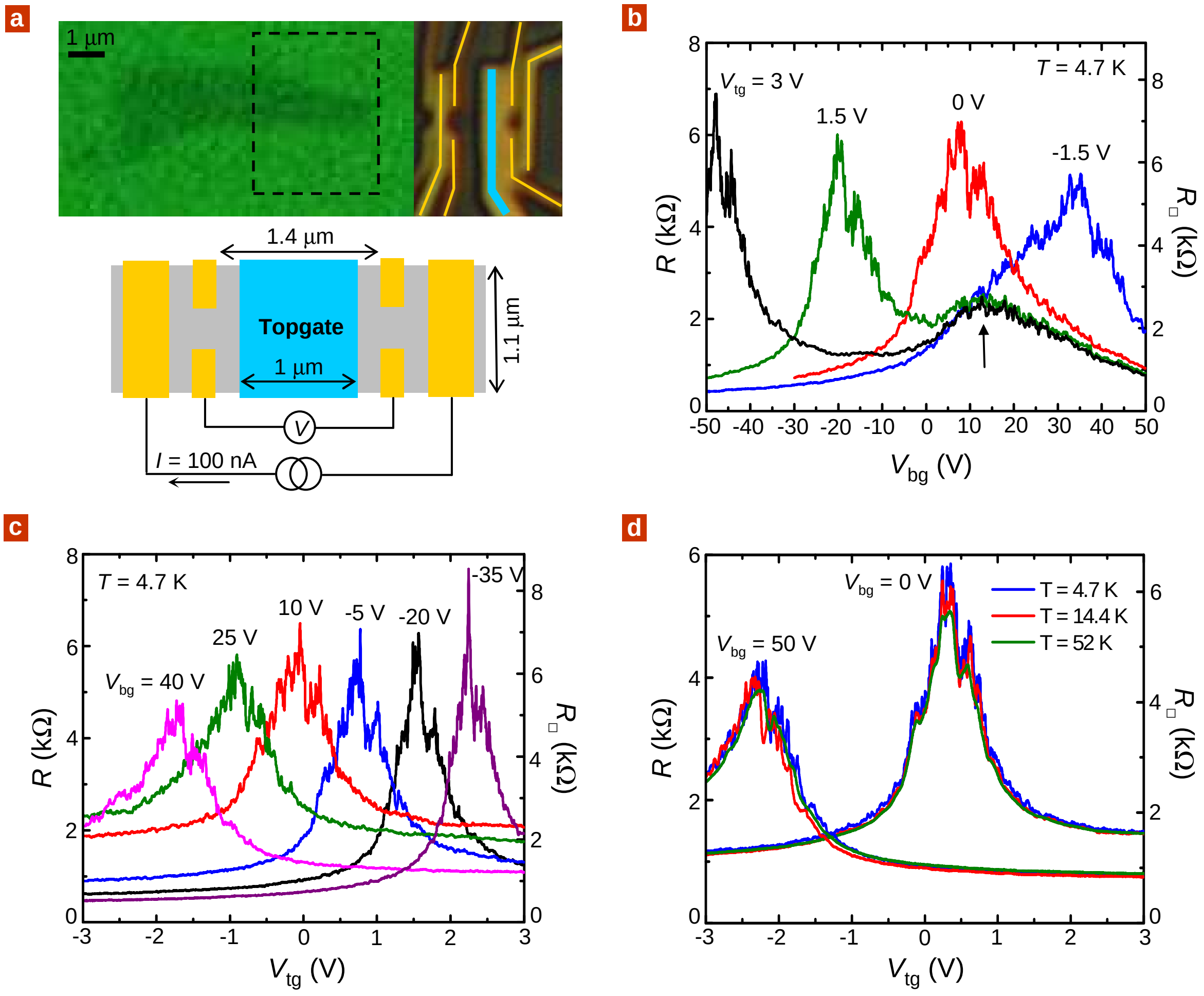

# Figure 3

**Oostinga *et al.***

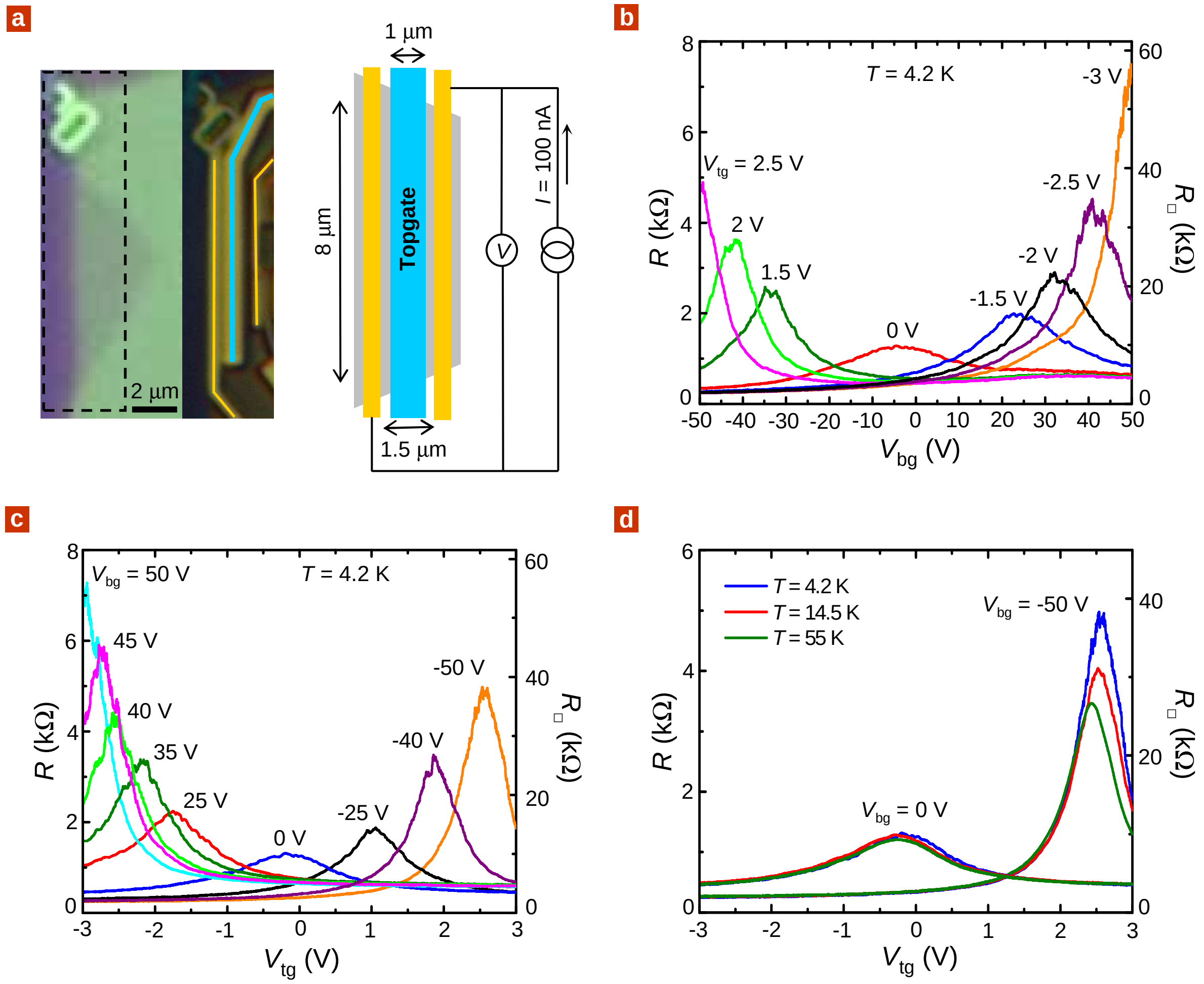

# Figure 4

**Oostinga *et al.***

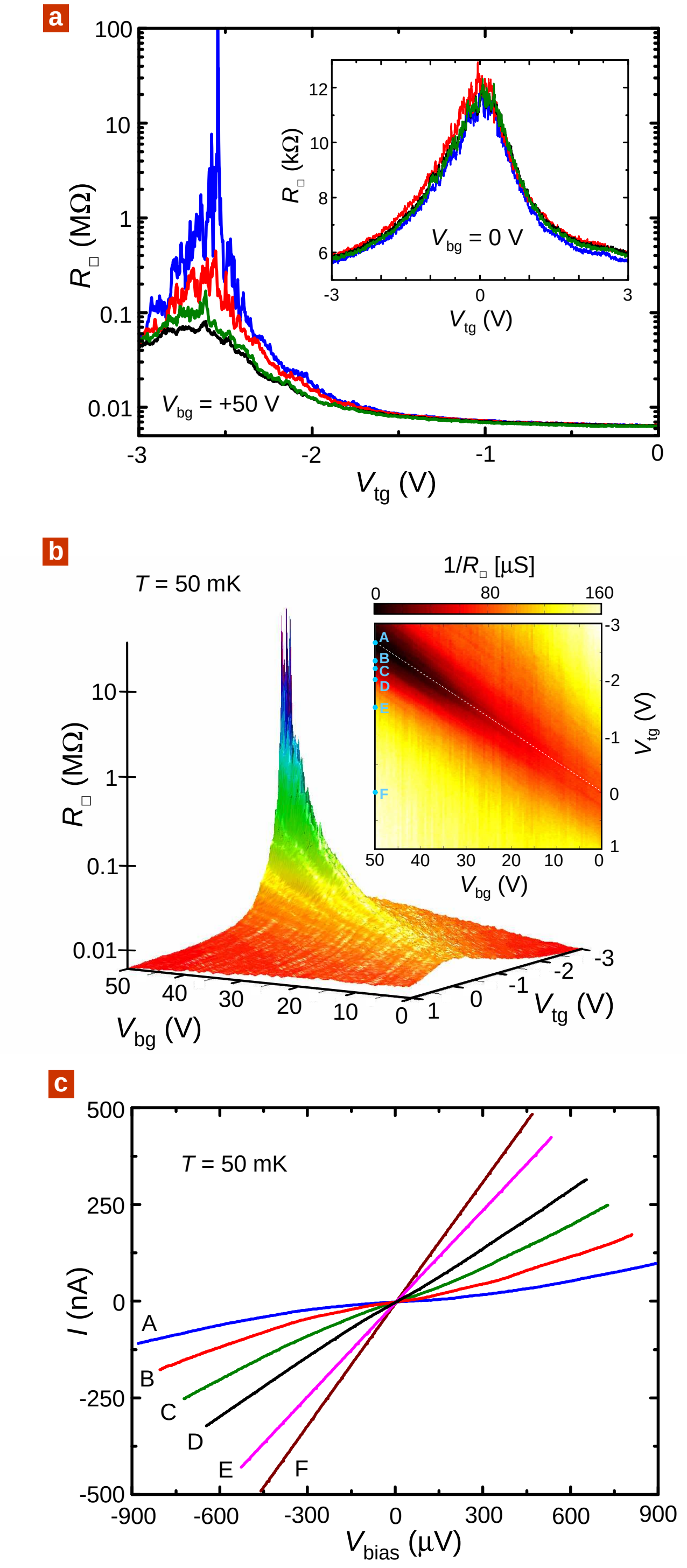

# Figure 5

**Oostinga *et al.***

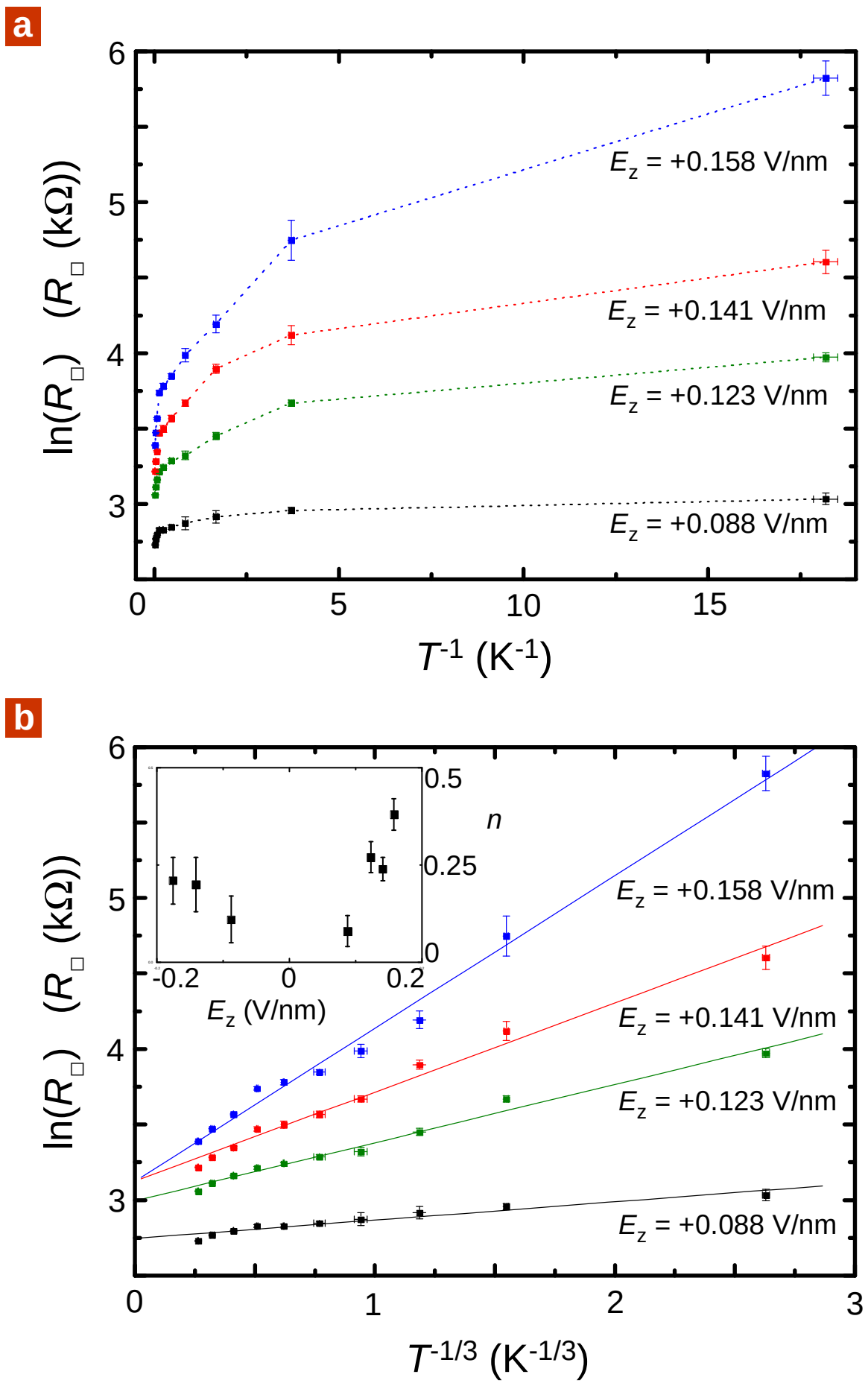